%% file: notebooksearch.tex
\documentclass[sigconf]{acmart}

\AtBeginDocument{%
  \providecommand\BibTeX{{%
    \normalfont B\kern-0.5em{\scshape i\kern-0.25em b}\kern-0.8em\TeX}}}

\setcopyright{none}
\copyrightyear{2021}
\acmYear{2021}
\acmDOI{10.1145/3411764.3445048}

\acmConference[CHI '21]{CHI '21: ACM CHI Conference on Human Factors in Computing Systems}{May 8--13, 2021}{Yokohama, Japan}
\acmBooktitle{CHI '21: ACM CHI Conference on Human Factors in Computing Systems, May 8--13, 2021, Yokohama, Japan}
\acmPrice{15.00}
\acmISBN{978-1-4503-8096-6/21/05}

\usepackage{graphicx}
\usepackage{soul}
\usepackage{tabularx}
\usepackage{xcolor}
\acmSubmissionID{4098}

\newcommand{\name}{NBSearch}
\newcommand{\visname}{NBLines}

\newcommand{\etal}{et~al.}
\newcommand{\eg}{e.g.}
\newcommand{\ie}{i.e.}

\newcommand{\q}[1]{\textit{``#1''}}

\def\markup{0}
\if\markup1
\newcommand{\rv}[1]{{\leavevmode\color{teal}#1}}
\else
\newcommand{\rv}[1]{#1}
\fi

\begin{document}

\title{\name: Semantic Search and Visual Exploration of Computational Notebooks}






\author{Xingjun Li}
\affiliation{%
  \institution{University of Waterloo}
  \streetaddress{200 University Ave W}
  \city{Waterloo}
  \state{ON}
  \postcode{N2L 3G1}
}
\email{x586li@uwaterloo.ca}

\author{Yuanxin Wang}
\affiliation{%
  \institution{University of Waterloo}
  \streetaddress{200 University Ave W}
  \city{Waterloo}
  \state{ON}
  \postcode{N2L 3G1}
}
\email{y2587wan@uwaterloo.ca}

\author{Hong Wang}
\affiliation{
    \institution{Uber Technologies}
    \city{San Francisco}
    \state{CA}
}
\email{hongw@uber.com}

\author{Yang Wang}
\affiliation{
    \institution{Uber Technologies}
    \city{San Francisco}
    \state{CA}
}
\email{gnavvy@gmail.com}

\author{Jian Zhao}
\affiliation{%
  \institution{University of Waterloo}
  \streetaddress{200 University Ave W}
  \city{Waterloo}
  \state{ON}
  \postcode{N2L 3G1}
}
\email{jianzhao@uwaterloo.ca}


\begin{abstract}
Code search is an important and frequent activity for developers using computational notebooks (\eg, Jupyter). 
The flexibility of notebooks brings challenges for effective code search, where classic search interfaces for traditional software code may be limited.
In this paper, we propose, \name{}, a novel system that supports semantic code search in notebook collections and interactive visual exploration of search results. 
\name{} leverages advanced machine learning models to enable natural language search queries and intuitive visualizations to present complicated intra- and inter-notebook relationships in the returned results.
We developed \name{} through an iterative participatory design process with two experts from a large software company. 
We evaluated the models with a series of experiments and the whole system with a controlled user study. 
The results indicate the feasibility of our analytical pipeline and the effectiveness of \name{} to support code search in large notebook collections.


\end{abstract}



\maketitle

\input{texfiles/1-introduction}
\input{texfiles/2-relatedwork}

\input{texfiles/3-design}
\input{texfiles/4-1-system}

\input{texfiles/4-2-system}

\input{texfiles/5-evaluation}
\input{texfiles/6-discussion}

\input{texfiles/7-conclusion}

\begin{acks}
This work is supported by the Natural Sciences and Engineering Research Council of Canada (NSERC) through the Discovery Grant. We would like to thank Chenqi Wan for providing the code for the MSA algorithm.
\end{acks}

\bibliographystyle{ACM-Reference-Format}
\bibliography{notebooksearch.bib}

 
\appendix

\input{texfiles/appendix.tex}

\end{document}

%% file: texfiles/1-introduction.tex
\section{Introduction}

\rv{Computational notebooks (\eg, Jupyter) combine code, documentation, and outputs (\eg, tables, charts, and images) within a single document, which provides expressive support for developers, such as adjusting code parameters to see results on the fly.
Searching examples, support, and documentation, such as which code APIs to use and how to use them, is one vital activity for developers using notebooks, as well as other development tools~\cite{head2019managing,rule2018exploration,chattopadhyay2020what,sadowski2015how}.} 
A notebook contains a collection of \emph{cells}, including \emph{markdown cells} for documentation and \emph{code cells} for scripts. 
Unlike traditional software code, these cells can be written and executed in any order, offering great convenience for developers. 

\rv{However, these unique features of notebooks increase the complexity of code and its management~\cite{head2019managing,wenskovitch2019albireo}, thus imposing challenges for code search.
Developers are often unknown about which APIs to use~\cite{sim2011how}, \eg, ``how can I plot a bar chart in Python?''
Semantic code search (\eg, based on module descriptions and data flows among APIs used by developers~\cite{mcmillan2012exemplar}) could address this issue by enabling natural language queries~\cite{wilson2010}; however, existing methods may not work appropriately because the structure of a notebook is nonlinear (with loosely-connected cells), unlike a traditional code module.} 
Cells in a notebook can be relatively independent and a group of cells can serve as a functional unit; meanwhile, they are interrelated, sharing data, variables, and function calls.     
We could treat each code cell as a basic search unit; however, there may not be enough semantic information (\eg, comments, docstrings, etc.) available to leverage. 
Further, to make sense of the search results, advanced user interfaces, rather than simple ranked lists, are needed~\cite{zhao2018flexible}, because of the complex relationships between returned cells and notebooks.  

To address these challenges, we propose a novel search interface, \emph{\name{}}, that provides effective code search in an extensive collection of notebooks and visual exploration of search results. 
\rv{As a first step, we design and develop \name{} as a standalone tool, rather than integrating into the Jupyter Notebook environment.}
We adopt the semantic search approach that has been demonstrated better performance and user experience~\cite{mcmillan2012exemplar}. 
By leveraging deep learning techniques, we enable this approach with individual notebook cells as basic search units, to accommodate the flexible nature of notebooks. 
Specifically, with a large notebook repository~\cite{ucsd2017corpus}, we train a \emph{translation model} to identify mappings between existing code and text descriptions (\eg, comments) in code cells. 
We then use the model to generate text descriptors for code without any comments, so each cell is associated with code and descriptors, which became our search database.
Next, we build a \emph{language model} that characterizes all the cell descriptors in a learned semantic space.
When a user initiates a query in natural language, \name{} utilizes the language model to process the query and retrieve search results (\ie, a collection of relevant cells from multiple notebooks) by searching in the semantic space of descriptors based on similarity and then retrieving associated code segments based on the code-descriptor pairs.
To facilitate the exploration of search results, we design a user interface that exhibits many aspects (\eg, variables used in code) in the search results. 
Specifically, it features an interactive visualization, \emph{\visname{}}, which reveals both intra- and inter-notebook relationships among cells. 

\name{} is developed through a user-centered, iterative, participatory design process with two senior developers at Uber (who are also the authors). 
Over the course of six months, we maintained regular meetings and consultations with the experts on multiple communication platforms (\eg, video conferencing, Slack, and emails), in which several prototypes were created and tested (\eg, sketches, low-fidelity, and high-fidelity prototypes).
We evaluated the machine learning models of \name{} by running comprehensive experiments to compare different model alternatives, using the public notebook repository~\cite{ucsd2017corpus}.
The results indicate that our analytical pipeline is promising and reveal deep insights into selecting models for building the pipeline. 
Moreover, we assessed the system as a whole by conducting user studies with 12 professional software engineers and data scientists at Uber by using their internal notebook collections. 
The results show that \name{} is useful and effective for supporting participants' code search tasks when programming with notebooks. 
\rv{The code associated with this paper can be accessed at \url{https://github.com/WatVis/NBSearch}.} 

In summary, our contributions in this paper include:
\begin{itemize}
    \item An analytical pipeline that enables the semantic search of code in computational notebook collections with the power of deep learning techniques;
    \item An interactive system, \name{}, that allows developers to conduct code search and browse search results enhanced by a visualization widget, \visname{}; and
    \item Experiments for both the system's analytical pipeline and user interface, which provides implications on selecting models for semantic search and designing search interface for computational notebooks.
\end{itemize}

%% file: texfiles/2-relatedwork.tex
\section{Background}

\subsection{Notebooks and Relevant Tools}
Computational notebooks (\eg, Jupyter Notebooks and RStudio) have recently emerged as a form of programming environments that support dynamic development. 
A notebook is broken down into \textit{cells} that appear in two primary forms, including code cells that contain executable scripts and markdown cells that contain formatted text (supplementing the comments in code cells). 
While a notebook is presented linearly, the order of executing the cells can be arbitrary, and each cell can be executed multiple times as well as edited between different runs, exhibiting a highly nonlinear programming pattern.

The dynamic nature of notebooks provides much freedom for developers, allowing them to experiment with different techniques, try out alternative methods, write temporary code, and conduct exploratory data analysis~\cite{kery2018story}.   
However, these features, at the same time, create challenges in code management, comprehension, and development. 
For example, messes in code may accrue during the process, and developers may lose track of their thought processes. 
To help track code history, several tools have been proposed, such as Variolite~\cite{kery2017variolite} and Verdant~\cite{kery2019effective}, which support fast versioning and visual exploration of history.
Code Gathering Tools~\cite{head2019managing} help find, clean, recover, and compare versions of code in Jupyter Notebooks.
Further, BURRITO~\cite{guo2012burrito} and TRACTUS~\cite{subramanian2020tractus} provide the provenance of analysis workflows by capturing and displaying code outputs, development activities, users' hypotheses, etc. 
More recently, the flexibility of notebook-style programming has been explored for providing code tutorials~\cite{head2020composing}.

\rv{Nevertheless, the above tools for notebooks assume developers produce the code, neglecting the support for code search activities that occur quite frequently in software development~\cite{sadowski2015how}. 
Developers have to use generic search engines (\eg, Google Search) or those built for traditional code repositories (\eg, SourceForge), whose effectiveness is unknown due to the uniqueness of notebooks.
In this paper, we focus on supporting code search for computational notebooks to improve developers' efficiency in identifying desired code samples and APIs to use.}

\subsection{Code Search and Visualization}
There exist several code search tools, in addition to generic search engines such as Google. 
Sim \etal~\cite{sim2011how} comprehensively compared Koders~\cite{wiki:koders}, Google, Google Code Search, Krugle~\cite{krugle}, and SourceForge~\cite{sourceforge}, with different sizes of search target and motivation for search. 
These search engines index code based on API keywords and method signatures, which may have limitations because code cells in notebooks are organized quite freely and the signatures are difficult to extract. 
Other approaches have been explored to provide more semantic features, allowing developers to search more effectively. 
Examplar~\cite{mcmillan2012exemplar} ranks search results using application descriptions, APIs used, and data-flows in code. 
CodeGenie~\cite{lemos2007codegenie} leverages a test-driven approach to search reusable code. 
Sourcerer~\cite{linstead2008sourcerer} uses fine-grained structural information from the code, enabling capabilities that go beyond keyword-based search.
\rv{However, none of the above search engines are designed to tailor searching code in computational notebooks, which have different characteristics compared to traditional code modules. 
For example, the code in notebooks is much messier, broken into small snippets (\ie, cells) that can be presented and executed in an arbitrary order. 
The analysis of code structures and semantics in existing tools relies on a clean, well-documented, and linearly-organized structure, which is often not available in notebooks. 
We approach this challenge by treating each cell as a search unit and identifying the relationships between cells within and across notebooks.}

On the other hand, approaches for software visualization have been developed to understand the functional and structural components of existing code. 
For example, UML (Unified Modeling Language) is a standard diagram for showing relationships between software components.
Hoffswell \etal~\cite{hoffswell2018augmenting} provided an in-situ visualization within the code to summarize variable values and distributions.
Based on a graph model, force-directed methods have been examined~\cite{seider2016visualizing}. 
Similarly, dependency graphs have been adopted to reveal connections between variables and functions~\cite{balmas2004displaying}.
Recently, Albireo~\cite{wenskovitch2019albireo} employs an interactive graph to visualize relationships of code and markdown cells in a notebook, while showing the information of variables, functions, and execution paths.  
TRACTUS~\cite{subramanian2020tractus} utilizes a tree structure to reveal the process of a hypothesis-driven analysis.
\rv{While the above techniques are useful for understanding code in a single notebook, they cannot be easily applied for visualizing multiple notebooks, which is the case for browsing search results.    
\name{} analyzes the relationships between cells within and between notebooks, presented in a novel visualization, \visname{}. It also reveals the connections among variables and functions used in a notebook, helping developers make sense of search results and discover serendipitous opportunities.}

\subsection{Search Results Exploration}
It is important for users to comprehend search results, especially when they do not know the exact answer; they may try to gain an understanding of a topic to retrieve more specific or related information for subsequent search actions~\cite{teevan2004perfect}.
Wilson \etal~\cite{wilson2010} advocated that the future web search interfaces should emphasize exploration. Users' search behaviors in web visualizations have also been explored~\cite{feng2018}.
There only exist a few works focusing on the sensemaking of code search results, many techniques have been proposed to visualize document and website searches or text.

One way of presenting search or recommended results is based on a linear ranked list like Google. 
TileBars~\cite{hearst1995tilebars} places a colored bar next to each list item to display document length and term frequency. 
Moreover, PubCloud~\cite{kuo2007tag} uses tag clouds to augment a ranked list. 
LineUp~\cite{gratzl2013lineupa} allows users to visually manipulate a multi-attributed ranked list. 
uRank~\cite{sciascio2016rank} provides on-the-fly search results refinement and reorganization as information needs evolve.

On the other hand, researchers have utilized a 2D space to present documents with various layout methods. 
One approach is to place items based on their attributes (or facets); for example, PivotPath~\cite{doerk2012pivotpaths} and PivotSlice~\cite{zhao2013interactive} display a publication corpus as a multi-attributed network and position each document interactively with a partition of a 2D canvas.
Another approach is projection-based methods.
Gomez-Nieto \etal~\cite{gomeznieto2014similarity} used an energy-based method to place text snippets of search results with minimal overlap. 
MOOCex~\cite{zhao2018flexible} visualizes a recommended list of videos on a 2D canvas using multidimensional scaling (MDS) with a specific similarity measure.
Moreover, space-filling techniques (\eg, treemaps and 2D tiles) have been used for browsing searching results~\cite{clarkson2009resultmaps,kleiman2015dynamicmaps}. 
Further, WebSearchViz~\cite{nguyen2006novel} and Topic-Relevance Map~\cite{peltonen2017topic} employ a circular layout, where the distance to the center represents a document's relevance, and the section of the circle denotes a specific topic or cluster. 
Likewise, RankSpiral~\cite{spoerri2004rankspiral} employs a spiral layout to render a ranked list by maximizing information density.

However, it is inadequate to directly apply the above methods in our scenario, because computational notebooks are different from text documents. 
For example, many of these techniques are based on topic analysis and text similarity, which cannot be extracted from code.  
\rv{Further, unlike documents or websites, the relationships (\eg, commonly-used variables and APIs) between code cells need to be revealed, but they have different distributions and characteristics compared to term relationships in documents.}

%% file: texfiles/3-design.tex
\section{Usage Scenario}
\label{sec:scenario}

\begin{figure*}[tb]
\centering
\includegraphics[width=\textwidth]{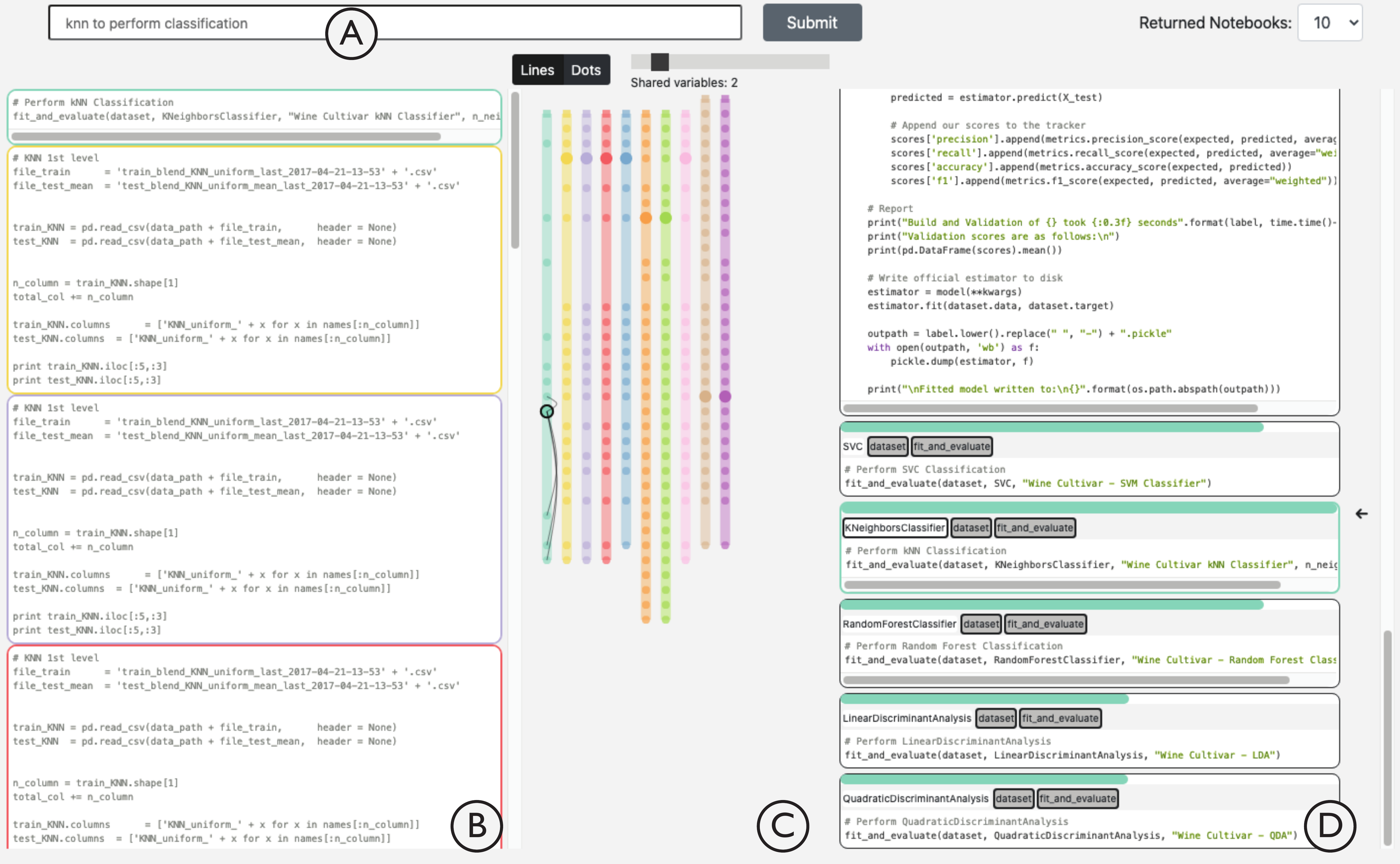}
\vspace{-7mm}
\caption{A developer is searching ``knn to perform classification'' with \name{} and exploring the results: (A) the Search Bar accepts a user's queries in natural language or code keywords; (B) the Search Result Panel displays a ranked list of relevant cells to the query; (C) the Notebook Relationship Panel visualizes intra- and inter-notebook relationships of cells based on the search result; and (D) the Notebook Panel shows the entire notebook containing the selected cell in the search results.}
\label{fig:usage}
\end{figure*}

In this section, we demonstrate the basic usage of \name{} with a motivating scenario. 
Detailed descriptions of the system design and functionalities will be introduced in the following sections.

Suppose that Tom is a software developer who is using Jupyter Notebooks to explore different solutions for solving a classification problem from his work. 
He is new to machine learning and has heard that the kNN model is a good start. 
But he does not know what toolkits or APIs to use or how to apply them.
Thus, Tom only has a vague goal in mind, so he types ``knn to perform classification'' in the search bar of the \name{} interface (\autoref{fig:usage}A). 

Within a second, a set of relevant notebook cells is returned from the search database, which is displayed as a ranked list on the Search Result Panel on the left (\autoref{fig:usage}B).
Each cell is colored uniquely to indicate the notebook it belongs to.
Simultaneously, the middle Notebook Relationship Panel (\autoref{fig:usage}C) displays \visname{}, a visualization of all the other cells within the same notebooks of the searched cells. 
Each notebook is visualized as a line, and each cell is shown as a dot, having the same color-coding in the Search Result Panel.  
Cells are also aligned based on their similarity and their order in the notebooks.

After briefly browsing the search results, Tom thinks the first returned item looks quite relevant. 
He clicks the cell, which triggers the Notebook Panel (\autoref{fig:usage}D) to appear, showing all the cells from that notebook.
\name{} automatically scrolls down to the searched cell, showing an arrow on the right of the Notebook Panel.

Tom then looks through the notebook and finds that it contains too many cells. 
Due to the complexity of the notebook (\eg, cells can be in arbitrary order), Tom has no idea how the searched cell is related to other cells in the same notebook, and thus does not know how to use the code in the cell in a meaningful context.
\name{} has a couple of features to support Tom to get the context. 
The colored bar on top of each cell indicates its similarity with respect to the search cell, and the tags below the bar show all the variables/functions used in a cell, where the ones sharing with the selected cell are highlighted in gray.
Further, from the black curves of the visualization on the Notebook Relationship Panel, Tom easily identifies all the cells having more than two overlapping variables/functions with the selected cell. 

With all the information, Tom finds that many other cells call the function \texttt{fit\_and\_evaluate}, similar to the searched cell. 
So, by following the arrows on the Notebook Relationship Panel, he decides to look into the cell that defines that function, which could be the code that he can actually reuse. 
Further, he discovers that other cells applying \texttt{fit\_and\_evaluate} use some other models, such as ``random forest classifier'' and ``linear discriminant analysis'', which are new concepts and methods that Tom has not been aware of(\ie, serendipitous findings). 
Thus, he also decides to look into these to see if they apply to his problem.

\section{\name{} Design}
\label{sec:design}

In this section, we introduce the design process of \name{} and the design goals that we distilled based on our consultation with two experts, followed by an overview of the system. 

\subsection{Design Process}

We employed a participatory design approach to develop \name{}, in which we actively involved two expert end-users over a six-month design process. 
Both experts work at Uber, which is a large IT company with more than 25K employees. One is a software engineer with more than three years of professional experience. The other is an engineering manager who has three years of development experience plus two years of management experience.  
They both use Jupyter Notebooks frequently every day to perform data analysis, conduct coding experiments, develop analytical models, etc. Moreover, both of them have once contributed to large open-sourced projects for Jupyter Notebooks.
\rv{Both experts are also authors of this paper, for their contributions to the deployment of \name{} using Uber's internal data and server as well as participant recruitment (see \autoref{sec:participants} and \autoref{sec:study_setup}). However, they were separated from the core research team for the design and development of \name{} at the University of Waterloo. }

\begin{figure}[tb]
    \centering
    \includegraphics[width=\linewidth]{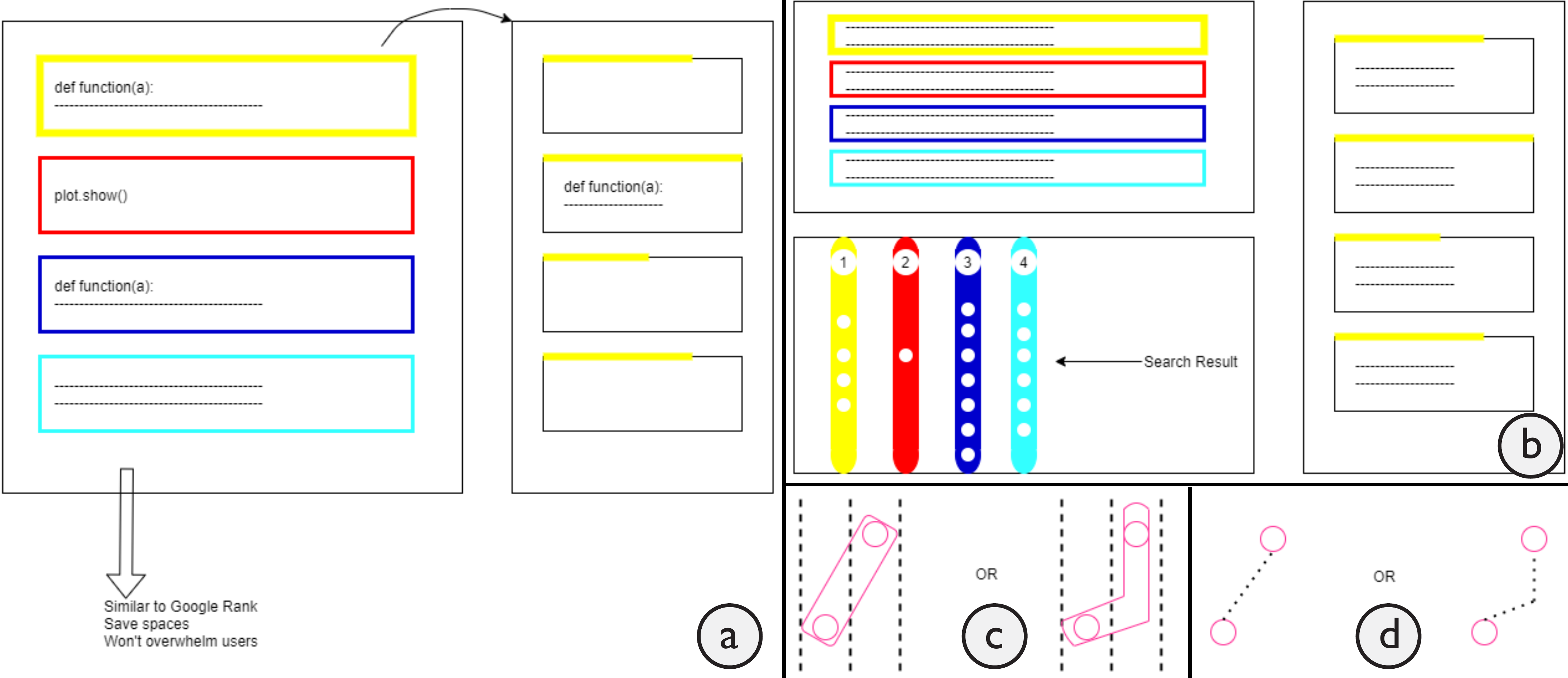}
    \caption{\rv{(a)(b) Early sketches for the \name{} user interface. (c)(d) Early visualization designs for showing the connections of multiple notebooks in the search results.}}
    \label{fig:sketches}
\end{figure}

Through the design process, we maintained bi-weekly meetings with the experts, during which they were presented with the latest ideas, designs, and prototypes for feedback on further improvements.  
We also set up a Slack channel to communicate irregularly for ad-hoc discussion, short consultation, and quick feedback. 
\rv{The whole design process consists of three stages: \textit{define \& ideate} (1.5 months), \textit{protype \& test} (3 months), and \textit{deploy \& evaluate} (1.5 months). 
During the first stage, we conducted a series of in-depth interview and brainstorming sessions with the experts. Together, we identified the common practices and seen problems of developers using Jupyter Notebooks in an enterprise setting. We thus iteratively distilled a set of design goals for guiding the development of code search systems with computational notebooks (see \autoref{sec:design_goals}). We also drew a blueprint for the system with some level of technical consideration (see \autoref{sec:overview}).
In the second stage, a series of prototypes were developed and refined based on our experts' feedback, including sketches, paper prototypes, analytical components with various machine learning models, and an interactive working prototype running on a public notebook corpus. During our meetings with the experts, we invited them to try out our prototypes and to comment on various aspects including visual design, interface layout, user interaction, search quality, etc. 
\autoref{fig:sketches} shows some early sketches and mockups for \name{}, which set the foundation of our final design shown in \autoref{fig:usage} (also see \autoref{sec:system_ui}). Our development of the backend analytical pipelines was also supported by a set of micro-benchmarks and experiments (see \autoref{sec:search_engine}).
For the third stage, with the help of our experts, we deployed a fully-functional prototype of \name{} on an internal notebook corpus from their company. We also carried out a user study to formally evaluate \name{} as a whole with professional data scientists and software engineers at Uber (see \autoref{sec:user_study}).   
}

\subsection{Design Goals} \label{sec:design_goals}

As introduced earlier, we distilled the following set of design goals from the first stage of our design process.

\rv{\textbf{G1: Retrieve relevant code segments with fuzzy search.}}
Notebooks are designed to facilitate on-the-fly programming, in which developers need to constantly write and refine code to try different methods~\cite{head2019managing}.  
Code search activities with vague goals occur more often in such exploratory scenarios, and developers usually do not know what exactly they want to use (\eg, specific APIs)~\cite{sim2011how}.
\rv{Especially, our experts mentioned: \q{For new employees, it is infeasible to glance through all the notebooks due to the sheer volume. They often don’t know which keyword to use. This is where contextual search comes in handy.}}
Thus, the system needs to go beyond the traditional keyword-based search by supporting fuzzy search based on semantics. That is, a developer should be able to query in natural language by describing roughly what they need.

\rv{\textbf{G2: Accommodate the flexibility of computational notebooks.}}
\rv{\q{Seasoned developers often document and publish their workflow by leveraging notebooks' dynamic nature for easy reproducibility and interpretability,} said by the experts.}
However, unlike traditional software code, each code cell in a notebook is relatively independent, and cells can be executed at an arbitrary order and for multiple times. 
This results in loosely connected code segments in notebooks and a highly nonlinear programming pattern for developers~\cite{head2019managing,subramanian2020tractus,wenskovitch2019albireo}. 
Classic code search by indexing functional modules may not be appropriate, because such concepts are unclear in notebooks. 
Hence, the system should be based on cells to adapt their flexibility \rv{and enable the discovery of \q{micro-components or steps in developers' workflows.}}

\textbf{G3: Allow for exploring intra- and inter-notebook relationships.}
Due to the complex structure of notebooks, the results returned by a search engine, which contain code cells from multiple notebooks, can be complicated. 
Representing the results with a classic ranked list may not be sufficient~\cite{zhao2018flexible}.  
\rv{Our experts also mentioned: \q{Existing platforms such as Google Search and StackOverflow do not help much because most of their software solutions are built in-house.}}  
For each cell in the search result, code cells from the same notebook can be relevant to provide a context of the usage of the scripts in that cell; however, the cells before or after that cell may not be necessarily related because of the dynamic nature of notebooks.
Further, given a specific cell of interest, code cells from other notebooks, including the cells returned in the search result and other cells co-existing in the associated notebooks (but not in the search result), can be useful for serendipitous discoveries.
Therefore, the system should provide an effective means for helping developers explore such intra- and inter-notebook relationships to make sense of the search results.

\subsection{System Overview} \label{sec:overview}

\begin{figure}[tb]
    \centering
    \includegraphics[width=\linewidth]{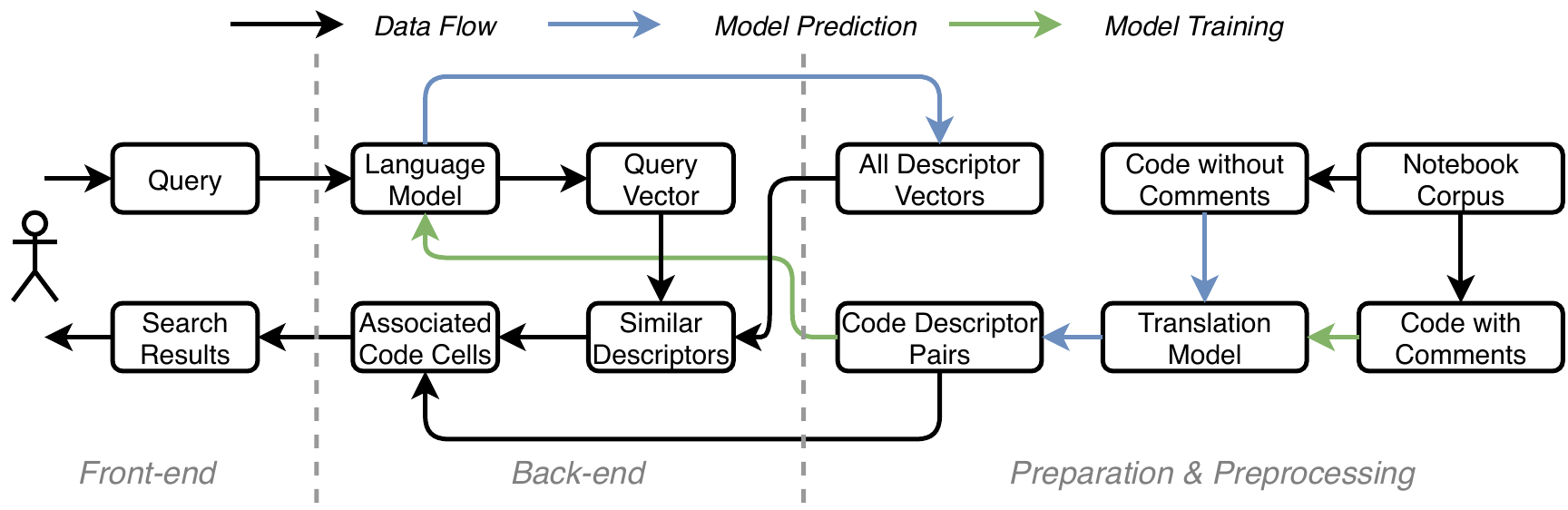}
    \vspace{-7mm}
    \caption{Overview of the \name{} system which consists of a front-end user interface and a back-end search engine that is built based on some preprocessing steps.}
    \label{fig:system}
\end{figure}

Following these design goals, we developed \name{} with the aforementioned iterative participatory design process with our experts. 
An overview of the system architecture is shown in \autoref{fig:system}. 
Details of each component will be introduced in the following sections. Here we provide a high-level description of the system.

First, a search database needs to be built from a notebook corpus, as illustrated in the preparation \& preprocessing step in \autoref{fig:system}.
For the notebook corpus, code cells were extracted and split into two sets: \textit{code with comments} and \textit{code without comments}. 
From the code cell with comments, a seq2seq \textit{translation model} (\eg, an LSTM recurrent neural network~\cite{lstm_model}) was trained to ``translate'' code scripts into corresponding comments, which we call text \textit{descriptors} of the code.
With the trained translation model, a completed database of \textit{code-descriptor pairs} of all the code cells in the corpus was generated by running the model on the code cells without comments. 
Then, all the text descriptors were sent to a \textit{language model} (\eg, Doc2Vec~\cite{doc2vec_theory}) to learn a semantic vector space of these texts.
Thus, each of the descriptors could be represented by a numerical vector, stored in a database of \textit{descriptor vectors}, where more similar descriptors have a closer distance in the vector space. 

With these trained models and generated databases, \name{} is able to handle a user's query and perform a semantic search of code in the notebook corpus.
As described in \autoref{sec:scenario}, when a user initiates a query in natural language on the search bar of the front-end interface, the back-end leverages the language model to convert this query into its vector form.
Then, it uses this \textit{query vector} to search in the formerly generated database of descriptor vectors and retrieve \textit{similar descriptors}. 
Based on these descriptors, the system further looks up in the database of code-descriptor pairs to obtain the \textit{associated code cells}. 
After, these code cells, along with other useful information (\eg, which notebooks they belong to, other cells within the same notebooks, variables/functions used in code), are sent to the front-end for search result presentation.

The semantic code search capability of \name{} does not mean to replace the classic code keyword-based search. 
Users can initiate a pure keyword search by using quotation marks on keywords (\eg, ``\texttt{plot.hist}''). 
This is useful when they know exactly what APIs to use and want to seek examples of how to use them.
The keyword-based search was implemented with the Okapi BM25 method~\cite{bm25} by indexing the scripts of notebook code cells, where exact keyword matches are found and ranked by their frequency in code cells.
However, \name{} does not currently support semantic and keyword-based searches simultaneously in one single query, which is an interesting future direction. 

%% file: texfiles/4-1-system.tex
\section{\name{} Search Engine} \label{sec:search_engine}

In this section, we introduce the development and evaluation of the back-end search engine of \name{}.

\subsection{Building Search Database} \label{sec:translation}
We first need a corpus of computational notebooks to construct a database to support code search. 
We obtained data from UCSD digital collection~\cite{ucsd2017corpus}, which contains 1.25M notebooks crawled from GitHub during 2017.  
As an initial attempt, in this paper, we used a sampled portion of the entire dataset for development and experimentation, which contains about 6K notebooks (around 10K code cells) as our corpus.
However, our method can be directly employed on the entire UCSD notebook collection with larger and stronger computing resources.

To support semantic code search (G1), at a high level, we want to build a database maintaining a one-to-one mapping between code segments and text descriptors in natural language. 
Therefore, we can process a natural language query to search similar text descriptors, and thus find the associated code segments performing the described functions, procedures, etc. 
The code itself cannot be a meaningful semantic descriptor, because it is significantly different from natural language.
Comments embedded in code are a straightforward source for constructing these descriptors, because they usually explain the underline semantic meaning of the code, among many other aspects such as control flows and desired outputs.

However, not every piece of code contains comments. 
Further, some comments are not in natural language (\eg, programmers comment out lines of code to compare different results). 
Thus, to utilize the code of the whole corpus, one approach is to generate the ``comments'' based on the code. 
Intuitively, we view it as a translation task in natural language processing, where we aim to build a model for translating programming language (\eg, Python code) into natural language (\eg, comments, or text descriptors), with the existing code and comments as the training data. 
With the trained model, we can generate descriptors from code that is without them, and thus build a complete dataset for all the code in the corpus (\autoref{fig:system}).

As we consider notebook cells as the basic search unit (G2), we processed the notebook corpus by cells. 
We first scanned the entire corpus and extracted code cells with comments. 
Then, we concatenated all the comments in a code cell to form the text descriptors for this cell and stored pairs of code and descriptors as our training set. 
We did not consider markdown cells, although they are potential semantic descriptors for the code in a notebook.
This is because it is challenging to identify which code cells are associated with which markdown cells.
A markdown cell may describe multiple code cells in a notebook (which can be the cells immediately after or some other cells), or just generally describe the settings of the entire notebook (\ie, associated with no code).
Due to this level of complexity, we decided not to use markdown cells since imprecise, noisy training data may affect the performance of the model. 
However, in the future, if there exists a high-quality labelled dataset showing the connections between markdowns and code in notebooks, our method can be easily applied and extended.       

\rv{As mentioned above, we formulated our problem as a translation task in machine learning, \ie, from Python code (programming language) to text descriptors (natural language). 
With recent advances in neural machine translation, we employed the \textit{seq2seq model} which is composed of an encoder and a decoder; both of them are recurrent neural networks (RNN)~\cite{rnn_knowledge}.
We considered four different state-of-the-art RNNs in this study, including: GRU (gated recurrent unit)~\cite{gru_model}, LSTM (long short-term memory)~\cite{lstm_model}, LSTM with Attention~\cite{attention_mechanism}, and BiLSTM (bidirectional LSTM)~\cite{bilstm_model}.
To assess their performance, we employed the BLEU (bilingual evaluation understudy) score~\cite{bleuscore}, which is a common measure of machine translation models. 
The results indicate that LSTM with Attention performed the worst, and the other three models performed similarly on our dataset. 
We thus manually browsed the results of GRU, LSTM and BiLSTM, and finally chose LSTM based on our observations and consultations with our experts. 
However, for deployment in practice, experiments might be required for applying the models on different datasets. 
Details about the data preprocessing, model training, and evaluation can be found in \autoref{apx:translation}.}

\subsection{Enabling Semantic Search} \label{sec:language}

Using the above trained translation model, we were able to generate descriptors for the code without sufficient comments. 
By combining the code cells with comments, we built a complete dataset of code-descriptors pairs for all the notebook code cells (\autoref{fig:system}).
However, having the descriptors for each cell is not enough to support semantic code search, as a user most likely input some queries that do not match the descriptors exactly.
We need to characterize the similarity between the queries and the available descriptors in semantics (G1). 
Therefore, we leveraged neural language models in natural language processing, which could represent words, phrases, or sentences with vectors in a latent space.  
These vectors can be stored in the database, based on which we could retrieve semantically similar descriptors from an input query by simply computing their cosine similarity (\autoref{fig:system}).

\rv{In this study, we considered two different language models: Doc2Vec~\cite{doc2vec_theory} and Sentence-BERT~\cite{sbert_theory}.} 
The goal of building the language model is to find descriptors that are semantically similar to a search query, however, this ``semantically similar'' criteria is very subjective. 
Unlike common language modeling tasks for English, there lacks ground truth or a standard benchmark to assess the performance of the models.
While most of the comments are written in English words, the usage patterns and distributions may be different. 
\rv{Inspired by the literature (\eg, \cite{sbert_theory}), we evaluated the two models using the downstream search tasks by conducting a survey study.
We carefully selected 10 queries, such as ``knn to perform classification'' and ``image manipulation.'' We then generated 10 sets of search results for each of the models. 
In our survey study, we asked participants to select which ones of the 10 items are relevant in the search result, and then to indicate which search result of the two models is better. 
Based on the results, we concluded that Doc2Vec is more suitable for the language model with our dataset. However, more experiments might be needed when the models are deployed on another notebook corpus.
Details about the model training, survey procedure, and results can be found in \autoref{apx:language}.}

%% file: texfiles/4-2-system.tex
\section{\name{} User Interface} \label{sec:system_ui}

With the aforementioned search engine, we could retrieve a set of code cells according to a natural language query initiated by a user. 
However, presenting the search results in a ranked list, like the traditional search engine, is not adequate to understand the complicated intra- and inter-notebook relationships between cells (G3). 
We designed a novel user interface for \name{} to allow developers to make sense of the search results interactively and visually, in order to accommodate the flexible and dynamic nature of computational notebooks. 

\subsection{Search Result Presentation}
As shown in \autoref{fig:usage}, the \name{} user interface consists of four components: (A) the Search Bar, (B) the Search Result Panel, (C) the Notebook Relationship Panel, and (D) the Notebook Panel. These panels of \name{} are interactively linked to many aspects, such as mouse hovering events. 
\rv{This interface was iterated several times with different layouts and design, as indicated in \autoref{fig:sketches}ab.}

After \name{} is launched, only the Search Bar is shown, and a developer can configure the number of notebook cells to return (\ie, top-$k$ results where $k$ is 10 by default) at the top-right corner of the UI. 
We restricted that only one cell from a notebook can be included in the top-$k$ results, allowing for extending the search scope a bit. 
This is because some code cells within a notebook are very similar, because a developer sometimes tries different methods in exploratory programming by simply changing one or two parameters of the code~\cite{wenskovitch2019albireo}.
If there is no such restriction, the search results may contain several very similar code cells. 
In fact, the back-end search engine of \name{} returns the whole notebooks containing the searched code cells to the front-end for display, and thus the information is not lost due to this restriction. This constraint can be turned off when needed.   

After processing a query, the Search Result Panel (\autoref{fig:usage}B) appears to display a ranked list of searched code cells.
Meanwhile, the Notebook Relationship Panel (\autoref{fig:usage}C) shows an interactive visualization---\visname{}---of the searched code cells and the notebooks they belong to.
The colors on both panels correspond to different notebooks; when there are more than 20 unique notebooks, the ones beyond the top 20 are colored in gray. 
Whenever a cell is clicked on the Search Result or Notebook Relationship Panels, the Notebook Panel (\autoref{fig:usage}D) shows the entire notebook containing that cell and jumps to that cell automatically. 
The colored bar on top of each cell indicates its similar with respect to the searched cell; so that the searched cell always has a full-length bar. 
Also, a series of tags indicate the variables and functions used in that cell, with dark gray tags showing the overlapping variables/functions between the selected cell and other cells. 
These features of the Notebook Panel help reveal the intra-notebook relationships.

\subsection{\visname{}}

\begin{figure*}[tb]
    \centering
    \includegraphics[width=\linewidth]{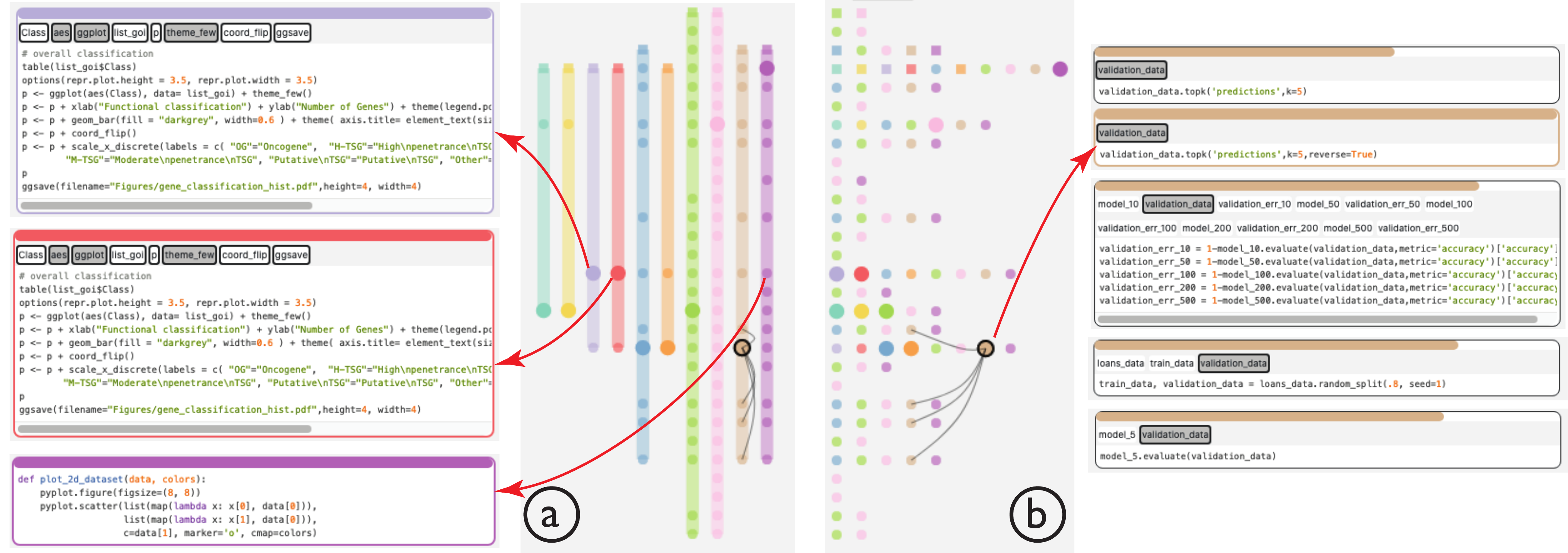}
    \vspace{-7mm}
    \caption{Visualizing the search results of query ``classification'' in \visname{}; (a) the lines view aligning cells (dots) of different notebooks (lines) based on similarity, and (b) the dots view using a similar alignment but shifting to the left to show the distribution of cells.}
    \label{fig:dot_line}
\end{figure*}

The \visname{} visualization displays both the intra- and inter-notebook cell relationships, allowing a developer to find connections between cells while maintaining a mental model of all notebooks. 
It contains two visual forms: the \textit{lines} view and the \textit{dots} view. 
In the lines view (\autoref{fig:dot_line}a), each notebook is visualized as a colored line, with cells represented as dots on the line, where the searched cell has a bigger size and a darker shade.
The visual design was inspired by EgoLines~\cite{Zhao2016Egocentric} that present relationships between actors in a temporal dynamic network.
The cells across multiple notebooks are aligned based on their similarity. 
That is, each row, across the lines, contains cells that similar to each other; at the same time, the order of the cells in each notebook (\ie, columns) is kept.
The order of the columns (or lines) is based on the search rank of the searched cell in that notebook, as indicated on Search Result Panel. 
The visualization reveals the inter-notebook relationship of cells, allowing a developer to look for similar cells across different notebooks, starting from the searched cell. 
This enables serendipitous findings in search results exploration.  

When a cell is selected, several black curves appear to indicate the cells sharing more than $n$ variables/functions within the same notebook.
The threshold $n$ can be adjusted interactively using a slider on top of the view. 
This reveals the intra-notebook relationship, allowing a developer to find relevant cells within a notebook. 
For example, in \autoref{fig:dot_line}b, the red arrow (on the right) indicates the cells linked by the black curves in the brown notebook all share the \texttt{validation\_data} variable.
It facilitates the sensemaking of a notebook's structure, because the cells can be out of order. 

The alignment of the notebook cells is very similar to the multiple sequence alignment (MSA) problem in computational biology, where several DNA or RNA sequences are aligned together to identify similar regions of nucleotide or amino acids in the sequences.
MSA is an NP-hard problem, and a number of approximation algorithms have been proposed. 
In particular, we employed a progressive multiple sequence alignment algorithm~\cite{multiseq} for its accuracy and speed. 
Such an approach progressively aligns multiple sequences by conducting pairwise sequence alignments in order and adjusting the global alignment.  
Our case is unlike the sequence alignment in computational biology, where the nucleotide or amino acids in different sequences need to be the same when aligned. 
We employed the cosine similarity between code cells to define the ``optimal'' alignment---having the smallest total similarity score for every pair of aligned cells in two sequences.   
The cell similarity can be easily computed because our search database contains the vector representation of all the cells. 
For example, in \autoref{fig:dot_line}a, the three red arrows (on the left) point to two exactly same (but from two different notebooks) cells (blue and red) and one similar cell (purple) on the same row, which all contains code about plotting.  

The dots view of \visname{} (\autoref{fig:dot_line}b) displays the same information as the lines view, using the same horizontal alignment and user interactions.
However, cells are pushed to the left and each column may contain cells from multiple notebooks.
This allows a developer to see the distribution of similar cells on all the rows. 
\rv{As shown in \autoref{fig:sketches}cd, we also experimented with different visual designs for connecting the cells across notebooks in the dots view during the design process. In the end, we removed them because our experts thought the visuals were too overwhelming.}

%% file: texfiles/5-evaluation.tex
\section{User Study} \label{sec:user_study}
We conducted a user study with a set of search tasks based on a real-world notebook corpus from Uber. 
Our specific goals of the study are to (S1) assess the usability of \name{} in supporting code search for notebooks, (S2) explore the potentials of \visname{} and other result exploration support, and (S3) understand users' behaviors in code search with \name{}.
\rv{We did not conduct a comparative study since there are no appropriate baselines. 
Our experts indicated that Google Search is the most commonly used for searching code in the public domain; however, the search corpus includes lots of other stuff than code. 
Our experts’ company uses an internal StackOverflow platform; however, it is mainly for question answering. 
Neither has the capability of searching at the cell level of notebooks or exploring the results. 
Thus, we decided to conduct a user study with the above goals that are exploratory in nature.}

\subsection{Participants} \label{sec:participants}
\rv{We recruited 12 participants, nine males and three females, aged 27--41 (four did not report it), through internal mailing lists at Uber with the help of our experts. 
They were all professional software engineers or data scientists with diverse backgrounds and from different engineering groups, covering a range of cases. Three of them have Bachelor's, six have Master's, and three have PhD degrees.} 
The two experts involved in our design process were excluded from the participants pool.
On a 1-7 Likert Scale, the medians of participants self-reported familiarity to Jupyter Notebooks and visualization techniques were 4.5 ($Mode=5$) and 4.5 ($Mode=3$), respectively (1 = ``no experience at all'' and 7 = ``using almost everyday'').
For the frequencies of conducting code search, six participants indicated they search code for 1-5 times daily, two for 6--10 times, two for 11--15 times, and two for over 20 times.

\subsection{Study Setup} \label{sec:study_setup}
We rebuilt our back-search engine by retraining the translation and language models based on an internal notebook corpus at Uber, in order to best capture participants' interests.
The corpus contains 5,583 notebooks with around 365K code cells, where around 95K cells have comments. 
These notebooks were created for manipulating data, training and testing machine learning models, generating visualization dashboards, and some other tasks. 
We followed the same process to build the search engine using the same sets of parameters, as described in \autoref{sec:search_engine}.
\rv{We retrained the four translation models and found that GRU performed the best on this dataset (see \autoref{apx:deployment}).} 
Based on our previous experience (see \autoref{sec:language}) and detailed observations of the results, we chose Doc2Vec as the language model for our search engine in the study because of its better and stabler performance.

\subsection{Procedure}
The study was conducted using a remote conferencing software, where participants used their own computers to access \name{} deployed on the company's internal server.
During the study, participants first completed a brief training session to get familiar with the \name{} interface by watching a short video. 

Then, they were asked to conduct three free-form code search tasks under different predetermined scenarios, for example, ``Imagine that you are writing some python code to \textit{plot your data with some charts}, but you have no clue. Now please use the search engine to see how others write relevant code and find desirable results.''
The search scenarios were carefully developed with our expert users who know the dataset and their daily activities very well. 
The goals were to prevent participants from searching stuff they are not interested or getting too few results due to the sparsity of the dataset.
At a high level, the three task scenarios were about: how to train machine learning models (T1), how to plot charts from data (T2), and how to use visualization tools (T3).
In each task, they were allowed to use either \name{}'s semantic or keyword-based search capabilities, or a combination of both.
There was no time limit for the search task, and participants were instructed to stop until they found what they needed in their minds. 
The order of the three tasks was counterbalanced across participants.

After, participants completed a post-study questionnaire about their impression of \name{}, followed by a short semi-structured interview.
Their search queries and interactions were logged, and the interview sessions were audio-recorded. 
The whole study lasted about 30 minutes for each participant.

\subsection{Results}

In the following, we report our results from the user study, including participants' performance in the search tasks, their questionnaire ratings, and their feedback to the system.

\subsubsection{Task Performance}

\begin{figure}[tb]
    \centering
    \includegraphics[width=\linewidth]{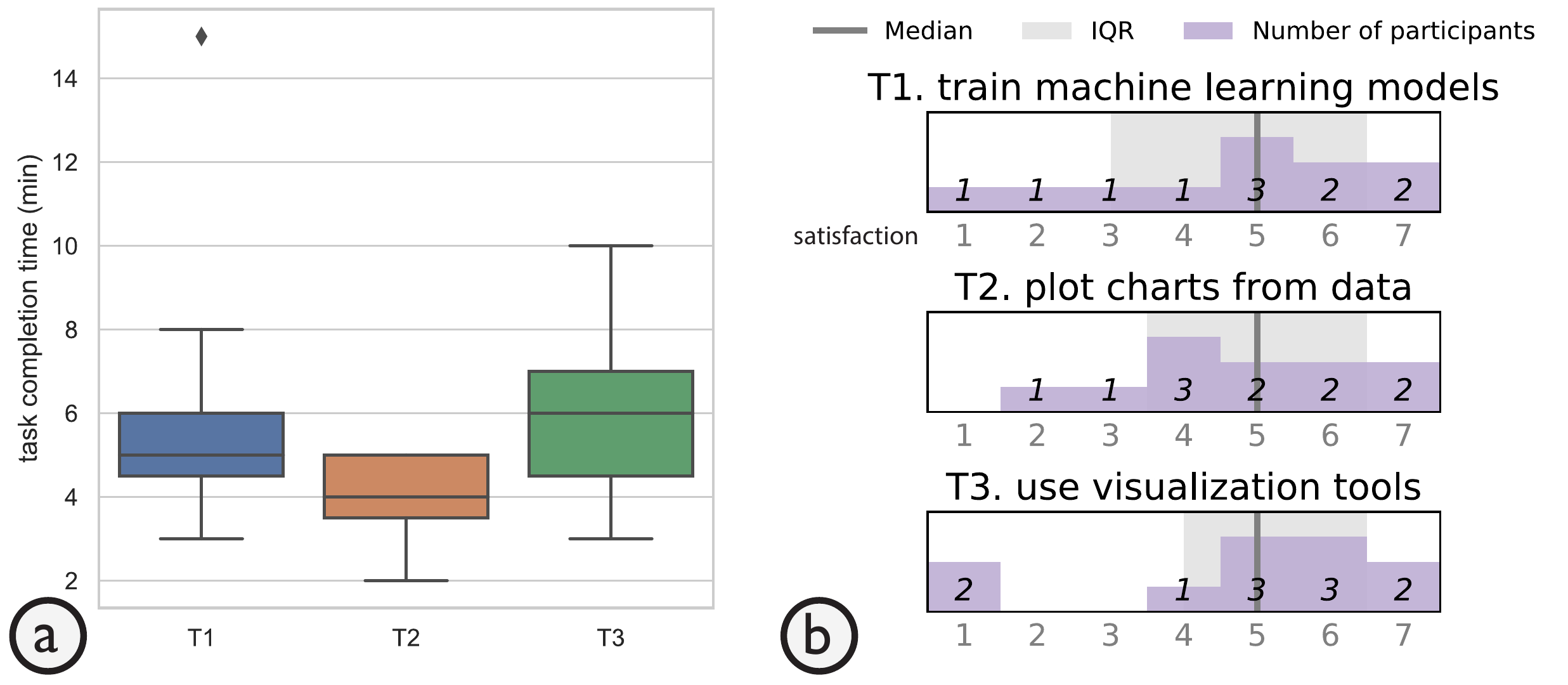}
    \vspace{-7mm}
    \caption{(a) Completion times of the three search tasks. (b) Participants' satisfaction level for the results (1=``very unsatisfied'' and 7=``very satisfied'').}
    \label{fig:search_time}
\end{figure}

\begin{table*}[tb]
    \centering
    \small
    \caption{Participants' search queries. Queries with quotation marks indicate the keyword-based search and those without indicate the default semantic search. ``$\rightarrow$'' indicates the sequence of conducting multiple search queries in one task.}
    \label{tab:search_queries}
    \vspace{-4mm}
    \begin{tabular}{cp{6.2in}}
        \toprule
        \textbf{T1} & regression model, Train $\rightarrow$ sklearn $\rightarrow$ ``sklearn'', Train $\rightarrow$ training, Train, Random forest $\rightarrow$ svd $\rightarrow$ bayes model $\rightarrow$ linear regression, model.fit $\rightarrow$ model.train, driver $\rightarrow$ sushi $\rightarrow$ san francisco, machine learning train $\rightarrow$ xgboost train, pipeline, Train, model train, model train
        \\ \midrule 
        \textbf{T2} & plot some charts, plot chart, ``plot'', plot, Plot $\rightarrow$ ``plot'', Histogram, plt.show $\rightarrow$ matplotlib plt plot $\rightarrow$ plt scatterplot, matplotlib, how to plot a chart $\rightarrow$ plot matplotlib, sns.plot, Chart, plot chart, chart plot
        \\ \midrule 
        \textbf{T3} & visualization, visualization, Pyplot $\rightarrow$ ``pyplot'' $\rightarrow$ visualization, Charts $\rightarrow$ bar chart $\rightarrow$ visualization $\rightarrow$ plot, visualization $\rightarrow$ plotly $\rightarrow$ manifold $\rightarrow$ ''manifold'', bucket $\rightarrow$ sampling $\rightarrow$ data frame std $\rightarrow$ standard deviation $\rightarrow$ ``mean'', df.filter $\rightarrow$ pandas df.filter $\rightarrow$ numpy.matrix.transpose, Sunburst $\rightarrow$ bar, data visualization, PaleteeTransformer $\rightarrow$ transformer, Utilization, michelangelo, pandas Matplotlib plt pd $\rightarrow$ plt pd $\rightarrow$ pd apply  
        \\ \bottomrule
    \end{tabular}    
\end{table*}

To investigate S1, we computed the average task completions times for the three search tasks: 5.67 ($\sigma=3.28$), 4 ($\sigma=1.04$), and 5.83 ($\sigma=1.85$) minutes, respectively (\autoref{fig:search_time}a). We can observe that participants were able to find the desired code within a couple of minutes. 
There is no obvious difference between different tasks in terms of the completion time, indicating that \name{} performed stably.

Moreover, on average, they conducted 1.83 ($\sigma=1.03$), 1.33 ($\sigma=0.65$), and 2.5 ($\sigma=0.65$) queries to find their targets for the three tasks, respectively.
Participants' self-reported satisfaction with the results found with \name{} is shown in \autoref{fig:search_time}b.
On a 7-point Likert scale (1=``very unsatisfied'' and 7=``very satisfied''), the medians of the satisfaction for the three tasks were all 5, with $IRQs=3.5$, 3, and 2.5, respectively. Thus, overall, \name{} could allow participants to find what they needed within a couple of queries with a relatively positive experience. 
However, there were a few cases where participants were not satisfied with the (partial) results. For example, one participant in T1 and two in T3 rated 1 (very unsatisfied); this was because they searched for three or more times before hitting their desirable targets. 
But note that as the search tasks are open-ended, there are no definitive right or wrong answers. One reason for the above lower ratings was that the results were limited by the code in the dataset that suited for participants' individual needs. 
As required by the tasks, they were able to find final satisfying results but some complained about the relatively long time of searching. No participant gave up in the middle of the tasks.

Further, to examine S3, we recorded the queries conducted by participants, as shown in \autoref{tab:search_queries}. 
All participants used the semantic search, and three used a combination of semantic and keyword-based searches. They also mentioned that previous search results helped narrow down the exact APIs to look up.
\rv{While participants were informed by both types of search, 83.33\% of the queries were purely based on semantic, 2.78\% purely based on keywords, and 13.89\% mixed (using both methods but in different queries). This indicates the tendency of participants' behaviors.} 
Computed for all the tasks together, 52.78\% of the search activities allowed participants to find their target within one query, 72.22\% within two queries, and 88.89\% within three queries. There was no case that participants conducted more than five queries (only one participant had five). 
This indicates that in general \name{} performed well in finding relevant code within the notebook collection.

\subsubsection{Questionnaire Responses}

\begin{figure*}[tb]
    \centering
    \includegraphics[width=\linewidth]{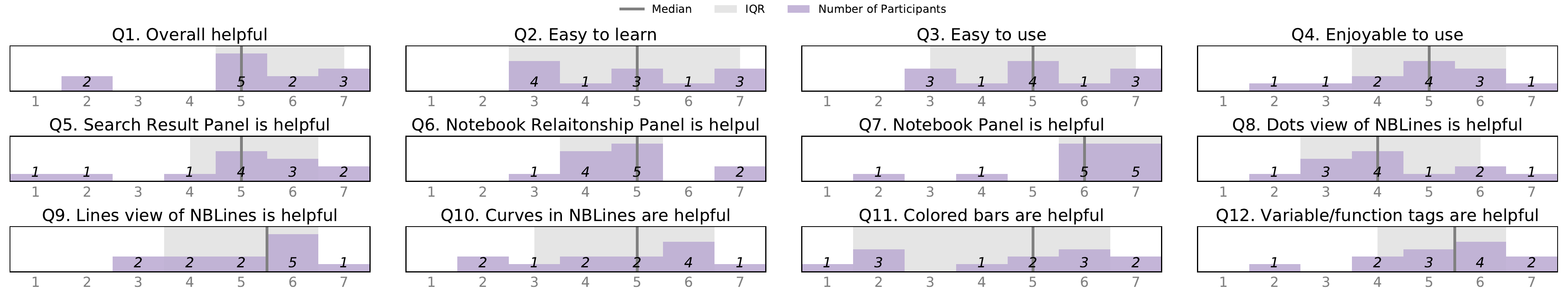}
    \vspace{-7mm}
    \caption{Participants' ratings on the post-study questionnaire (1=``strongly disagree'' and 7=``strongly agree'').}
    \label{fig:questionnaire}
\end{figure*}

\autoref{fig:questionnaire} shows participants' ratings on their impression with \name{} on the post-study questionnaire, revealing insights into S1 and S2.
Overall, \name{} was perceived helpful (Q1), easy to learn (Q2), easy to use (Q3), and enjoyable to use (Q4), all having a median rating of 5. However, the $IRQs$ for easy to learn and easy to use were larger, 4.5 and 4, respectively. This might be because participants have very different backgrounds in visualization, affecting their acceptance to \visname. 
Participants also felt that the three panels were helpful (Q5-7), all with a median of 5 or more. Two participants rated the Search Result Panel (\autoref{fig:system}B) lower because it took them longer (more than three queries) to get the desired results. Two participants applauded for the Notebook Relationship Panel (\autoref{fig:system}C), giving the highest rating. The Notebook Panel (\autoref{fig:system}D) was perceived very useful, too. 
For \visname{}, participants had divergent opinions on the dots view (Q8), with a median of 4 ($IQR=3.5$), but they overall preferred the lines view (Q9), with a median of 5.5 ($IQR=3$).  
The curves connecting shared variables/functions were perceived useful in general (Q10), but several mentioned that the view became too messy when there were many cells connected.
For the visual design of the Notebook Panel, there was a large variance in the usefulness of the colored bars on top of each cell (Q11), but participants thought the tags for variables and functions were useful (Q12).

\subsubsection{Qualitative Feedback}
\rv{Our interview mainly focused on collecting participants' feedback on all the aspects of \name{} in order to assess S1-3 qualitatively. 
In general, participants thought \name{}'s visual encodings and interface were intuitive, helpful, and clear. They also like the colors: \q{It’s comfortable for me to see this interface.}-P11. 
In the following, we report our results based on the design goals in \autoref{sec:design}.} 

\rv{\textbf{Keyword-Based and Semantic Search.} 
Supporting the semantic search in addition to the keyword-based approach is one of our key goals (G1). 
Eleven out of 12 participants agreed that having both options would improve their efficiency and appreciated the ability to search in natural language. P10 commented: \q{For advanced users, they know what they are searching, so they would prefer keyword-based more [...] I only knew one library can do visualization, but after using semantic search, I find other libraries are useful.}
They also demanded the two options to be merged in the future. 
But P2 was not convinced about semantic search and thought that \q{the results of the keyword-based search is more predictable.} 
For all the five cases involving keyword search (\autoref{tab:search_queries}), there were two for P2, two for P4, and one for P5. However, P2 and P4 said that the previous semantic searches in the same task inspired their keywords. 

\textbf{Cell-Based Search and Presentation.}
To accommodate the dynamic nature of notebooks, we built \name{} with code cells as the central unit (G2).
We found that participants had different opinions on how many cells should be displayed in the search result. Three of them thought it should be below 10, and the others thought it should be between 10 and 25. Seven participants felt that the default setting (\ie, 10 items) was appropriate, as P11 said: \q{Usually people [who] use Google Search don't check the second page.} 
Some returned code cells were very short; while they were relevant, it was difficult for participants to understand without having enough context. Thus, merging of related cells in the search results may be developed in the future to allow for a broader search coverage.
Regarding the workflow, P10 suggested switching the position of the Notebook Panel (\autoref{fig:usage}D) with the Notebook Relationship Panel (\autoref{fig:usage}C), \q{because usually I will firstly browse if the notebook contains good results, and then I will check relationships of cells.}

\textbf{Intra- and Inter-Notebook Exploration.}
To facilitate sensemaking of search results, we designed \visname{} (\autoref{fig:dot_line}) to support exploring intra- and inter-notebook relationships (G3). 
Participants appreciated \visname{} for its usefulness of \q{visual guidance for finding results.}-P9. Also, P6 said that \q{I think it's useful because I can realize the position that search result locates in a notebook, even if sometimes I still need to read the whole notebook's code.} 
The black curves showing the overlapping variables and functions among cells were applauded by some participants, such as helping \q{trace back of the code.}-P2. P4 added: \q{One of the features that I love is the link. I would like to see which cells are using the same variables as the current cell when I am doing code search.}
While liking it, P5 hoped to \q{see some details about the notebook in the middle view (Notebook Relationship Panel).}
By adjusting the threshold of overlapping variables/functions, participants were able to refine their search results easily, as commented by P7: \q{After changing it a little bit, I realized it can help me narrow down our search results.} P11 especially liked \q{the idea to use the filter,} so that she could clearly see the connections between cells of interests.

Several participants found the dots view (\autoref{fig:dot_line}b) was confusing, \q{because cells with the same color are not on the same column.}-P1.
They also demanded more time to get familiar with the visual design. P3 commented: \q{I’m used to a traditional code search task; you search, and here are the results; you click it, and you see it.}   
P4 wanted to reduce the complexity of the black curves in \visname: \q{Ability to focus on variables I selected can help me find results very fast.}

Some visual cues are designed to augment the Notebook Panel (\autoref{fig:usage}D) for supporting a detailed exploration of a selected notebook. 
Regarding the colored bar for showing the cell similarity, half of the participants said that it was a nice-to-have feature but they did not rely on it. 
P7 said that it might be \q{a little distracting,} and P3 \q{later found they are representing correlations.}
The rest of the participants thought it was useful, especially P1: \q{I highly rely on that to find results.} Additionally, P4 thought \q{the progress bars are robust and reliable.}
The tags showing the variables/functions were appreciated, and P6 suggested to make them more obvious.
} 

%% file: texfiles/6-discussion.tex
\section{Discussion}
Here, we discuss the limitations of \name{} and our study, as well as the possible avenues for extending and generalizing our design.

\subsection{Limitations and Future Work}
While our study results indicate that \name{} can facilitate developers with code search during exploratory programming in large computational notebook repositories, it still has several limitations. We discuss them below and introduce potential solutions.

First, we did not involve markdown cells in building our search database of code-descriptor pairs, because it is hard to precisely identify which code cells are described by which markdown cell. 
Albireo~\cite{wenskovitch2019albireo} uses a naive method by simply associating a markdown cell with the code cell after.
However, this oversimplifies the scenarios, given that notebooks are extremely flexible; sometimes a markdown cell may not correspond to any code in a notebook, and sometimes a markdown cell may describe a group of cells that are not necessarily the ones following it. 
It is very challenging to detect the relationships between code and markdowns using unsupervised methods, which needs a considerable amount of labelled data available in the future. 

Second, currently \name{} only enables semantic code search at the cell level, as the flexibility and organization of notebooks are based on cells; however, future extensions can be made to support search at sub-cell and multi-cell levels.  
For the sub-cell level, we can leverage the location of comments and the functional scope of code. 
For example, if a comment is inline with the start of a for-loop, we can pair all the code within the scope of the loop with this comment (or descriptors), similar for functions and statements.   
By enhancing the cell-level code-descriptor pairs with such information, the search engine is able to return finer or coarser-grained results based on a user's query; and the analytical pipeline of \name{} can naturally adapt to this new data without changes.

Third, the front-end interface and back-end models of \name{} can be enhanced. 
In the controlled study, several participants mentioned that they spent quite some time to get familiar with the system interface, especially the \visname, while they thought it was useful to explore the search results. They also thought the dots view was confusing. Future improvements may include removing the dots view and providing on-demand tutorials for the lines view. 
\rv{Also, \name{} currently works as a standalone tool, but future development of \name{} within the Jypyter Notebook eco-system will provide much convenience to users. This will also require setting up a public server for the search APIs.}
For the search engine, there were a couple of participants who were not very satisfied with the search results for the first few attempts.  
Trying other models or training them on a larger repository could resolve this issue; however, this may require humongous data with which the models can gradually become better, such as Google Search.  

Finally, the evaluation of \name{} can be further enhanced. 
Additional and larger testing datasets need to be used to thoroughly assess the performance of the translation model; and a large Mechanical Turk study, in a similar form to our survey study, may be conducted to deeply examine the quality of the language model. 
While we designed \name{} through an iterative participatory design process with two experts and evaluated the system with professionals in the user study, a longer-term deployment is necessary for investigating the usefulness of \name{} in practice. 
\rv{While there are currently no appropriate baselines, a comparative study between \name{} and its base version (\eg, without \visname{} and other visual cues) would provide deeper insights into the impact of the user interface.} 
We leave these studies for future work.

\subsection{Generalization and Extension}
We propose a general architecture to support semantic code search as illustrated in \autoref{fig:system}. 
It is flexible enough to incorporate other translation and language models when more advanced ones are available in natural language processing research. 
Our test scripts also provide insights into which models to use for particular data in practice.
\rv{As the business of each company may differ, retraining may be needed on their dataset for better retrieval; this is because different datasets may have different distributions and patterns of code usage.
Our experts encountered no difficulty in retraining and deploying \name{} with their dataset, which indicates the generalizability of our computational flow. 
We validated \name{} with a large pool of professionals from Uber, supporting that our design can be generalized to various cases and needs. However, further validations in different cases need to be conducted.}  

Moreover, the \visname{} visualization, equipped with a notebook alignment algorithm, provides an easy means of browsing the intra- and inter-notebook relationships of cells. It allows users to better understand the notebooks' structures and discover serendipitous findings. 
This can be extended to show sub-cell level relationships when needed, \eg, by representing the functional unit of code as circles and cells as circle groups. 
It is useful for analyzing a collection of computational notebooks at a finer level, as sometimes code cells can be quite long depending on a developer's habits.  
The visualization itself is general enough to apply on other problems involving the comparison of multiple sequences, such as clickstream data in web analytics.

%% file: texfiles/7-conclusion.tex
\section{Conclusion}
We have presented \name{}, a tool that facilitates code search within large computational notebook repositories. 
\name{} leverages machine learning techniques to support semantic search in natural language queries, thus allowing developers to vaguely specify their needs during exploratory programming.
A series of experiments were conducted to compare different models for the back-end search engine.
The tool also features an interactive front-end, including a visualization \visname, to present the search results in an intuitive manner, revealing the complicated relationships among cells in multiple notebooks. 
The development of \name{} was through a user-centered, iterative, participatory design process with two professional developers at Uber. 
Further, the tool was evaluated with developers and data scientists through a user study with an internal dataset at the same company. 
Participants' feedback to \name{} reveals its usefulness and effectiveness in supporting code search for computational notebooks.

%% file: texfiles/appendix.tex
\rv{
\section{Search Engine Development Details}

\subsection{Translation Model} \label{apx:translation}
As introduced in \autoref{sec:translation}, we formulated our problem as a translation task in machine learning, \ie, from Python code (programming language) to text descriptors (natural language).
We processed the corpus by concatenating all the comments in each code cell.
Comments with non-English words were ignored for simplicity, and also comments formed by code where developers simply commented out lines of code during programming. 
This is because these comments are not considered as meaningful descriptors in natural language.
In total, we had 27,423 pairs of code and descriptors in our training set, leaving
77,935 code cells to generate descriptors. 

\subsubsection{Model Training}
We employed the seq2seq model with a common encoder-decoder architecture with RNNs~\cite{rnn_knowledge}.
The encoder captures the context of an input sequence into a hidden state vector, and the decoder produces an output sequence from this hidden state vector. 
In machine translation tasks, the current practice also includes an embedding layer and a batch normalization layer before sending sequences of words (or tokens) to the RNNs, for a faster and stabler performance~\cite{batch_norm}.

We considered four different RNNs: GRU (gated recurrent unit)~\cite{gru_model}, LSTM (long short-term memory)~\cite{lstm_model}, LSTM with Attention~\cite{attention_mechanism}, and BiLSTM (bidirectional LSTM)~\cite{bilstm_model}.
For all the four models, we set the size of the embedding layers to 300. 
The input and output sequences were tokenized, where the input language (\ie, code) had a vocabulary size of 8,002 and the output language (\ie, comments) had a vocabulary size of 4,502. 
We used 1-layer RNNs with the GRU/ LSTM having 300 units; however, for BiLSTM, it was doubled because there are two directions to learn.
We trained these models with a batch size of 120 for 30 epochs. 
Within all the samples obtained from the preprocessing stage, we used 88\% of the data for training and 12\% for validation.

\subsubsection{Model Evaluation}
We used the BLEU score~\cite{bleuscore} to assess the performance of our models.
It is a metric (from 0 to 1) for evaluating a generated sentence (\ie, the produced text descriptors) to a reference sentence (\ie, the original text descriptors associated with the code), where a score of 1 indicates a perfect match.
The BLUE score is usually computed based on $n$ consecutive words/tokens (\ie, $n$-grams) in the sentences.
A perfect score is not possible in practice as a translation would have to match the reference exactly, which is not even possible by human translators. 
In general, a BLUE score over 0.5 would indicate that the translation model performs very well and significantly less post-editing is required to achieve publishable quality~\cite{google_bleu}. 

\begin{table}[tb]
\centering
\small
\caption{Results of the experiments on our public dataset for comparing the performance of the four translation models: GRU, LSTM, BiLSTM, and LSTM with Attention.}
\label{tab:bluescores}
\vspace{-4mm}
\begin{tabular}{ccccc} 
 \toprule
  & \textbf{GRU} & \textbf{LSTM} & \textbf{BiLSTM} & \textbf{LSTM w/ Att.} \\  
 \midrule
 \textbf{1-gram} & 0.57495 & 0.57607 & 0.58360 & 0.07025 \\ 
 \textbf{2-gram} & 0.58355 & 0.58127 & 0.59012 & 0.61298 \\
 \textbf{3-gram} & 0.58090 & 0.58145 & 0.59142 & 0.74337 \\
 \textbf{4-gram} & 0.59440 & 0.59422 & 0.60393 & 0.74951 \\
 \textbf{Cumulative Avg.} & 0.56372 & 0.56377 & 0.57301 & 0.47570 \\  
 \bottomrule
\end{tabular}
\end{table}

\autoref{tab:bluescores} shows the BLUE scores for 1- to 4-grams and the cumulative averages for the four models.
For the results, we can see that while LSTM with Attention performed well in the cases of 2- to 4-grams, its accumulative average was the lowest. 
On the other hand, the performances of GRU, LSTM and BiLSTM were similar and stable, resulting in higher cumulative averages than that of LSTM with Attention.
But it is difficult to conclude with the best model among the three; further, human subjective opinions may differ from the computed scores.
We thus manually browsed the results and finally chose LSTM based on our observations.

\subsection{Language Model} \label{apx:language}
\begin{figure*}
    \centering
    \includegraphics[width=\linewidth]{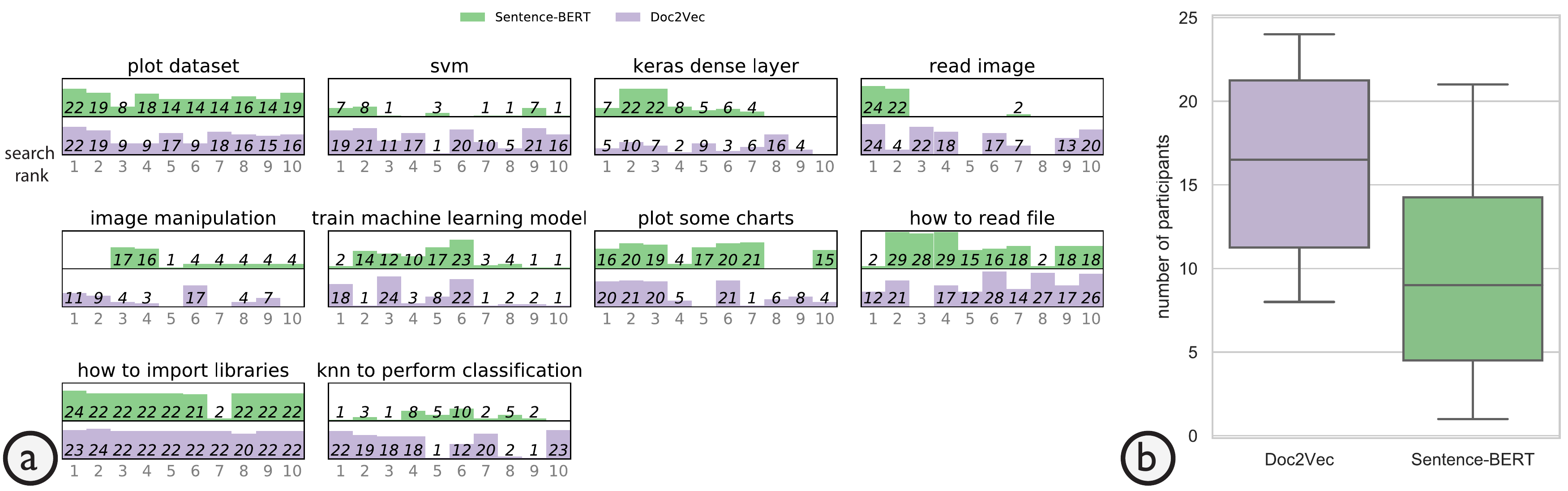}
    \vspace{-7mm}
    \caption{Results of the survey study on our public dataset for comparing the performance of the two language models: Doc2Vec and Sentence-BERT. (a) Distributions of participants' selections of the relevant items in the top-10 returned results for each query. (b) Boxplots of the numbers of participants who thought either of the two search results more relevant for each query.}
    \label{fig:survey_results}
\end{figure*}

As introduced in \autoref{sec:language}, we employed a language model to enable the semantic search of code based on the existing comments and generated ones with the translation model.

\subsubsection{Model Training}
We considered two different language models: Doc2Vec~\cite{doc2vec_theory} and Sentence-BERT~\cite{sbert_theory}.
Doc2Vec is an unsupervised learning model, which requires no labelled data. 
We thus trained our Doc2Vec model using all the descriptors from the database built in the last step.  
We set the latent dimension to 300 and the number of epochs as 40 using the APIs in the Gensim library~\cite{gensim_lib}.
The training of Sentence-BERT requires a large number of labelled sentence pairs that are similar in semantic meanings, which could not be easily produced in our cases. 
Fortunately, there are several pre-trained Sentence-BERT models that have demonstrated superb performance on many NLP tasks~\cite{bertlib}. 
In this study, we chose the \texttt{bert-base-uncased} model for its speed and performance, which has 12 layers, 768 hidden units, 12 heads, and 110M parameters, pretrained on lower-cased English text.

\subsubsection{Model Evaluation}
As described earlier, we conducted a survey study to evaluation the performance of the two language models.
We carefully selected 10 queries and generated 10 sets of search results for each of the models. 
These queries represent some common questions asked by developers in exploratory programming, based on our consultation with the two experts.
Each result contained the top-10 most similar code cells that we retrieved by computing the cosine similarity of the query and their descriptors.
In the study, for each query of each model, we asked participants to select which ones of the 10 items are relevant in the search result. They also needed to indicate which search result of the two models is better. 
The order of the 10 queries and the order of the two models were randomized in the survey study. 

We distributed our survey via mailing lists and social media at a local university. Finally, we collected 32 responses; however, some of the participants did not finish all the 10 questions. 
All participants have a computer science major. The medians of their self-reported familiarity to visualization tools (\eg, Matplotlib), machine learning tools (\eg, Keras), and Python are 4 ($Mode=4$), 4 ($Mode=5$), and 5 ($Mode=6$), respectively, on a 1-7 Likert scale (1 = ``very unfamiliar'' and 7 = ``very familiar''). 
As \autoref{fig:survey_results}b indicates, overall, participants thought Doc2Vec generated more relevant results than Sentence-BERT. 
As shown in \autoref{fig:survey_results}a, roughly, there are similar numbers of relevant items chosen by participants. 
For some queries such as ``plot dataset,'' ``how to read file,'' and ``how to import libraries,'' most of the items are voted relevant for both models.
Besides the number of relevant items selected for each query, the more top-ranked items are chosen, the better the model's performance is.
For most queries, especially ``svm,'' ``train machine learning model,'' and ``knn to perform classification,'' more participants chose items with smaller search ranks for Doc2Vec.
Further, for all the queries, there are more top-3 items chosen by participants for Doc2Vec, compared to Sentence-BERT. 

\subsection{Deployment Details} \label{apx:deployment}

\begin{table}[tb]
    \centering
    \small
    \caption{Results of the experiments on Uber internal dataset for comparing the performance of the four translation models: GRU, LSTM, BiLSTM, and LSTM with Attention.}
    \label{tab:bluescores-deploy}
    \vspace{-4mm}
    \begin{tabular}{ccccc} 
     \toprule
      & \textbf{GRU} & \textbf{LSTM} & \textbf{BiLSTM} & \textbf{LSTM w/ Att.} \\  
     \midrule
     \textbf{1-gram} & 0.62410 & 0.62203 & 0.61912 & 0.26917 \\ 
     \textbf{2-gram} & 0.55517 & 0.55542 & 0.55353 & 0.21200 \\
     \textbf{3-gram} & 0.44067 & 0.44257 & 0.44106 & 0.11814 \\
     \textbf{4-gram} & 0.34854 & 0.34851 & 0.34934 & 0.05638 \\
     \textbf{Cumulative Avg.} & 0.36927 & 0.36912 & 0.36981 & 0.06936 \\  
     \bottomrule
    \end{tabular}
\end{table}

We deployed \name{} with the internal framework from our experts' company, so that we could evaluate \name{} with professional developers on tasks and datasets that are more meaningful to them. 
The internal dataset contains 5,583 notebooks with around 365K code cells (around 95K cells with comments).
These notebooks were written by software engineers and data scientists from different groups at Uber, and they often have usages of internal libraries. 
We retrained the models using these notebooks and evaluated the models by comparing the BLEU scores and search results (\autoref{tab:bluescores-deploy}). 
From the BLEU scores, we can see that GRU, LSTM, and BiLSTM performed very similar, aligned with our experimental results with the public dataset. While the cumulative BLEU scores are lower than those in \autoref{tab:bluescores}, the results still reached the quality of ``good translations'' \cite{google_bleu}.  
We then manually examined the results of GRU, LSTM, and BiLSTM from a set of search queries via consultations with our experts, and finally selected GRU for the quality of content in results. 
We then deployed the \name{} user interface with this model. The deployment into the internal platform has flask applications hosted in docker images. One of the difficulties we encountered was to resolve the dependency conflicts between the library versions used in \name{} and the standard library presets in the docker image.
}